\documentclass[10pt,twocolumn]{article}

\usepackage{arxivstyle}

\providecommand{\ket}[1]{\left| #1 \right\rangle}
\providecommand{\bra}[1]{\left\langle #1 \right|}

\renewcommand{\vec}[1]{\boldsymbol{#1}}

\makeatletter
\newcommand{\doi}{doi:~\begingroup\catcode`\_=12\relax\doi@aux}
\newcommand{\doi@aux}[1]{\href{https://doi.org/#1}{\nolinkurl{#1}}\endgroup}
\makeatother

\makeatletter
\newcommand{\defval}[2]{\expandafter\def\csname val@#1\endcsname{#2}}
\newcommand{\val}[1]{\@ifundefined{val@#1}{\textcolor{red}{[\detokenize{#1}]}}{\csname val@#1\endcsname}}
\makeatother

\defval{totalCycles}{30.0}
\defval{totalMs}{187}
\defval{fftCycles}{13.6}
\defval{fftShare}{45\%}
\defval{timeShareMin}{1.7\%}
\defval{timeShareMax}{9.4\%}
\defval{inferenceCycles}{1,345}
\defval{optAdam}{36\%}
\defval{feasAdam}{90\%}
\defval{gapAdam}{0.103}
\defval{cyclesAdam}{9.8}
\defval{f1Adam}{0.637}
\defval{aucAdam}{0.793}
\defval{accAdam}{0.542}
\defval{optSpsa}{20\%}
\defval{feasSpsa}{100\%}
\defval{gapSpsa}{0.159}
\defval{cyclesSpsa}{7.8}
\defval{f1Spsa}{0.611}
\defval{aucSpsa}{0.779}
\defval{accSpsa}{0.521}
\defval{optCobyla}{32\%}
\defval{feasCobyla}{100\%}
\defval{gapCobyla}{0.114}
\defval{cyclesCobyla}{11.5}
\defval{f1Cobyla}{0.636}
\defval{aucCobyla}{0.787}
\defval{accCobyla}{0.541}
\defval{optRandom}{0\%}
\defval{feasRandom}{36\%}
\defval{gapRandom}{0.988}
\defval{cyclesRandom}{17.9}
\defval{f1Random}{0.691}
\defval{aucRandom}{0.845}
\defval{accRandom}{0.613}
\defval{optExact}{100\%}
\defval{feasExact}{100\%}
\defval{gapExact}{0.000}
\defval{cyclesExact}{14.2}
\defval{f1Exact}{0.775}
\defval{aucExact}{0.841}
\defval{accExact}{0.670}
\defval{optFree}{33\%}
\defval{feasFree}{33\%}
\defval{gapFree}{-0.061}
\defval{cyclesFree}{14.8}
\defval{f1Free}{0.737}
\defval{aucFree}{0.834}
\defval{accFree}{0.633}
\defval{pAnswerAdam}{98\%}
\defval{optAdamR1}{0\%}
\defval{gapAdamR1}{0.250}
\defval{optAdamR2}{100\%}
\defval{gapAdamR2}{0.000}
\defval{optAdamR3}{7\%}
\defval{gapAdamR3}{0.059}
\defval{pAnswerSpsa}{33\%}
\defval{optSpsaR1}{0\%}
\defval{gapSpsaR1}{0.329}
\defval{optSpsaR2}{60\%}
\defval{gapSpsaR2}{0.021}
\defval{optSpsaR3}{0\%}
\defval{gapSpsaR3}{0.127}
\defval{pAnswerCobyla}{37\%}
\defval{optCobylaR1}{0\%}
\defval{gapCobylaR1}{0.223}
\defval{optCobylaR2}{97\%}
\defval{gapCobylaR2}{0.002}
\defval{optCobylaR3}{0\%}
\defval{gapCobylaR3}{0.117}
\defval{vqeRuns}{90}
\defval{vqeSeeds}{30}
\defval{randomDraws}{1000}
\defval{budget}{18}
\defval{budgetMs}{112}
\defval{feasibleShare}{36\%}
\defval{budgetBinds}{2}
\defval{infeasibleBelow}{0}
\defval{f1Range}{0.00--0.99}
\defval{vqeSpsaShots}{$6.2 \times 10^{6}$}
\defval{feasAdamR1}{97\%}
\defval{feasAdamR2}{100\%}
\defval{feasAdamR3}{73\%}
\defval{feasSpsaR1}{100\%}
\defval{feasSpsaR2}{100\%}
\defval{feasSpsaR3}{100\%}
\defval{feasCobylaR1}{100\%}
\defval{feasCobylaR2}{100\%}
\defval{feasCobylaR3}{100\%}
\defval{f1RandomFeas}{0.672}
\defval{cyclesRandomFeas}{15.5}
\defval{f1AdamMiss}{0.525}
\defval{cyclesAdamMiss}{11.0}
\defval{gapAdamMiss}{0.160}
\defval{f1ExactBind}{0.740}
\defval{cyclesExactBind}{17.9}
\defval{relevanceRange}{0.25--0.82}
\defval{randomFftShare}{94\%}
\defval{f1RandomFft}{0.700}
\defval{aucRandomFft}{0.851}
\defval{cyclesRandomFft}{18.7}
\defval{f1RandomNoFft}{0.547}
\defval{aucRandomNoFft}{0.751}
\defval{cyclesRandomNoFft}{5.7}
\defval{vqeFftShare}{32\%}
\defval{vqeAdamSteps}{1181}
\defval{vqeSecondsCobyla}{55.4}
\defval{vqeSecondsSpsa}{28.6}
\defval{vqeSeconds}{22.7}
\defval{ocAccAdamDec}{0.636}
\defval{ocF1AdamDec}{0.742}
\defval{ocAucAdamDec}{0.817}
\defval{ocFaAdamDec}{24\%}
\defval{ocRecallAdamDec}{62\%}
\defval{ocGapAdamDec}{0.587}
\defval{ocBestAdamDec}{0\%}
\defval{ocAccAdamSrch}{0.673}
\defval{ocF1AdamSrch}{0.777}
\defval{ocAucAdamSrch}{0.839}
\defval{ocFaAdamSrch}{32\%}
\defval{ocRecallAdamSrch}{67\%}
\defval{ocGapAdamSrch}{0.013}
\defval{ocBestAdamSrch}{0\%}
\defval{ocAccSpsaDec}{0.587}
\defval{ocF1SpsaDec}{0.699}
\defval{ocAucSpsaDec}{0.840}
\defval{ocFaSpsaDec}{18\%}
\defval{ocRecallSpsaDec}{55\%}
\defval{ocGapSpsaDec}{0.947}
\defval{ocBestSpsaDec}{0\%}
\defval{ocAccSpsaSrch}{0.673}
\defval{ocF1SpsaSrch}{0.777}
\defval{ocAucSpsaSrch}{0.836}
\defval{ocFaSpsaSrch}{33\%}
\defval{ocRecallSpsaSrch}{67\%}
\defval{ocGapSpsaSrch}{0.006}
\defval{ocBestSpsaSrch}{0\%}
\defval{ocAccRandDec}{0.652}
\defval{ocF1RandDec}{0.750}
\defval{ocAucRandDec}{0.722}
\defval{ocFaRandDec}{33\%}
\defval{ocRecallRandDec}{65\%}
\defval{ocGapRandDec}{0.994}
\defval{ocBestRandDec}{0\%}
\defval{ocAccRandSrch}{0.673}
\defval{ocF1RandSrch}{0.777}
\defval{ocAucRandSrch}{0.837}
\defval{ocFaRandSrch}{33\%}
\defval{ocRecallRandSrch}{67\%}
\defval{ocGapRandSrch}{0.006}
\defval{ocBestRandSrch}{2\%}
\defval{ocAccSmoDec}{0.675}
\defval{ocF1SmoDec}{0.777}
\defval{ocAucSmoDec}{0.847}
\defval{ocFaSmoDec}{29\%}
\defval{ocRecallSmoDec}{67\%}
\defval{ocGapSmoDec}{0.001}
\defval{ocBestSmoDec}{0\%}
\defval{ocAccSmoSrch}{0.675}
\defval{ocF1SmoSrch}{0.778}
\defval{ocAucSmoSrch}{0.836}
\defval{ocFaSmoSrch}{33\%}
\defval{ocRecallSmoSrch}{68\%}
\defval{ocGapSmoSrch}{0.000}
\defval{ocBestSmoSrch}{67\%}
\defval{ocAucRoundMin}{0.56}
\defval{ocAucRoundMax}{1.00}
\defval{ocAucRoundMid}{0.96}
\defval{ocFaRoundMax}{97\%}
\defval{ocFaRoundMin}{0\%}
\defval{ocRecallRoundMin}{58\%}
\defval{ocRecallRoundMax}{84\%}
\defval{thresholdBa}{0.71}
\defval{ocRuns}{15}
\defval{ocRandom}{100}
\defval{fixedAgree}{99.9\%}
\defval{ocFeatures}{round 1: impulse\_factor, margin\_factor, skewness, band\_energy, spectral\_spread; round 2: rms, shape\_factor, clearance\_factor, margin\_factor, skewness; round 3: impulse\_factor, margin\_factor, skewness, band\_energy, spectral\_spread}
\defval{ocSpsaShots}{$1.8 \times 10^{7}$}
\defval{ocAdamSteps}{847}
\defval{ocSpsaIters}{3,000}
\defval{ocSpsaEvals}{6,020}
\defval{ocHoursAdam}{0.48}
\defval{ocHoursSpsa}{1.18}
\defval{ocAccSmoCont}{0.673}
\defval{ocF1SmoCont}{0.778}
\defval{ocAucSmoCont}{0.841}
\defval{smoSupport}{40--43}
\defval{smoBounded}{23--26}
\defval{smoFree}{16--20}
\defval{smoLbGap}{0.005}
\defval{smoRoundGapMax}{0.003}
\defval{smoRoundBest}{2}
\defval{ocRawGapPenalty}{0.92}
\defval{ocRawGapOffset}{0.59}
\defval{ocRawGapSpsa}{0.95}
\defval{ocRawGapUntrained}{0.93}
\defval{nJobs}{12}
\defval{parallelism}{1.9}
\defval{runHours}{2.5}

\makeatletter
\newcommand{\gentable}[2]{\IfFileExists{#1.tex}{\input{#1.tex} }%
  {\textcolor{red}{[\detokenize{#1}: run paper/values.py]} \\}}
\makeatother

\newcommand{\resultfigure}[2]{%
  \IfFileExists{#2}{\includegraphics[width=#1]{#2}}%
  {\fbox{\parbox[c][3.5cm][c]{0.95#1}{\centering\textcolor{red}{Pending: \texttt{\detokenize{#2}}\\(results/full/report/)}}}}}

\runningtitle{Predictive Maintenance at the Edge: Anomaly Detection Trained with Quantum Optimization}

\begin{document}

\sptitle{Preprint}

\title{Predictive Maintenance at the Edge: An Embedded Anomaly Detection Pipeline Trained with Quantum Optimization}

\author{Filipe Chagas Ferraz}
\affil{Department of Computing, Federal University of S\~ao Carlos, S\~ao Carlos, SP, Brazil}

\maketitle

\begin{abstract}
This article presents a predictive-maintenance pipeline that detects anomalies in bearings from vibration signals, adapted for deployment at the edge on low-cost microcontrollers, and that uses quantum optimization in two steps of its training. The pipeline consists of a step of vibration feature extraction and a step of classification with a one-class support vector machine (OCSVM). The features are selected by minimum redundancy and maximum relevance (mRMR) with the variational quantum eigensolver (VQE), under a limit on computing time. The OCSVM is adapted for inference at the edge without floating-point operations and is trained by a variational quantum algorithm with Pauli-correlation encoding (PCE). In small-scale tests on a simulator, with the Paderborn bearing dataset, the VQE found the optimal selection in \val{optAdam} of the runs with Adam, and PCE chose better support vectors than chance, until a classical local search made the two equal. A classical baseline, the usual OCSVM trained by SMO with its weights rounded, matched them without any search.
\end{abstract}

% Keywords: machine learning; predictive maintenance; quantum optimization;
% variational quantum eigensolver; Pauli-correlation encoding; anomaly
% detection; edge computing; support vector machines; feature selection.

%==============================================================================
% INTRODUCTION
%==============================================================================

\chapteri{I}ndustry depends on machines that wear out and can fail unexpectedly if the proper maintenance is not carried out. Maintenance can be preventive, done periodically; corrective, done after a breakdown; or predictive, done when a condition-monitoring system flags an anomaly early, possibly before it turns into a breakdown. The development of technologies that detect anomalies in machines from sensor data is an active topic called \emph{predictive maintenance} (PdM) \cite{carvalho2019}. In rotating machines, and in bearings in particular, an effective approach to detecting anomalies is to use machine-learning pipelines that classify vibration signals, because the characteristics of the vibration change when there are defects.

From a data-engineering point of view, a conventional way of deploying a PdM pipeline is to install the sensor on the machine, send the recorded data to the cloud, run the machine-learning pipeline in the cloud, and show the output on a dashboard. Another way, which can be cheaper, is to run the pipeline at the edge, that is, on an embedded system physically close to the sensor, with no processing in the cloud. This second approach becomes even more attractive when the prediction must be made in real time, under hard latency limits. In this case, however, the pipeline must be optimized for the limitations of the embedded hardware used. One such limitation is the high cost of floating-point operations. Low-cost microcontrollers, such as the ESP32-C6 and the STM32F103, have no floating-point unit (FPU), and floating-point operations take more clock cycles than integer or fixed-point ones. In FPGAs, which can serve as powerful prediction accelerators at the edge, floating-point operations take up more configurable logic blocks than integer and fixed-point ones.

\begin{figure*}
\centering
\includegraphics[width=\textwidth]{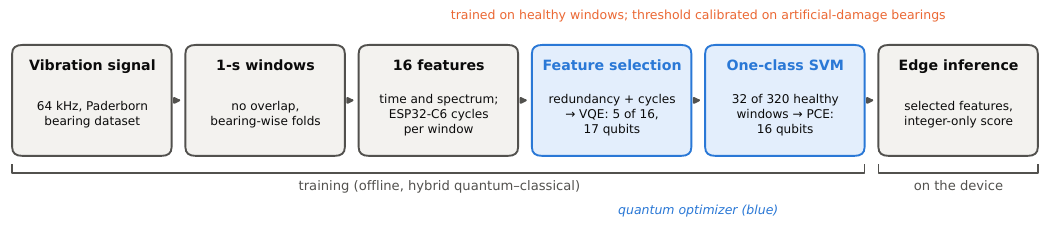}\\[2pt]
{\small (a)}\\[8pt]
\begin{minipage}[b]{0.36\textwidth}\centering
\includegraphics[width=\linewidth]{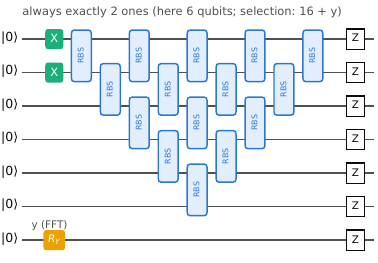}\\[2pt]
{\small (b)}
\end{minipage}\hspace{0.08\textwidth}
\begin{minipage}[b]{0.32\textwidth}\centering
\includegraphics[width=\linewidth]{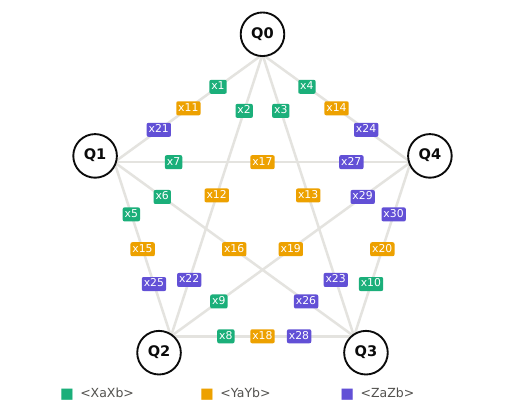}\\[2pt]
{\small (c)}
\end{minipage}\\[8pt]
\includegraphics[width=\textwidth]{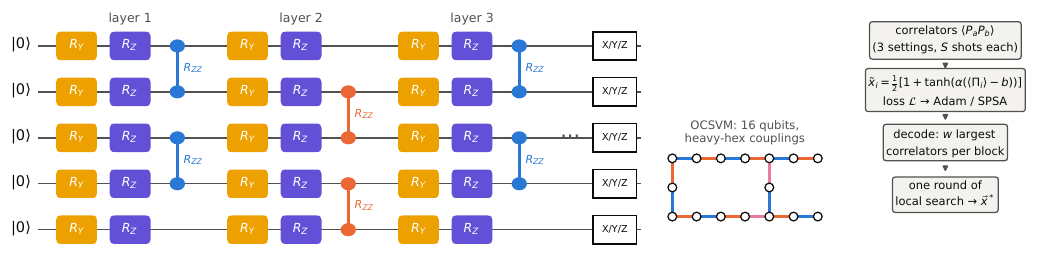}\\[2pt]
{\small (d)}
\caption{(a) The detector: during training, feature selection (VQE) and the OCSVM (PCE) are solved by quantum optimizers; the OCSVM learns from healthy windows only, and the relevance of the features and the decision threshold use artificial-damage bearings; on the microcontroller, inference uses integers only. (b) The VQE that preserves the number of ones, shown with 6 qubits and 2 ones: $X$ gates prepare the ones, a pyramid of RBS gates mixes them without changing their number, and a free qubit $y$, with one $R_Y$ rotation, stands for the FFT. (c) The PCE encoding: each edge $(a, b)$ of 5 qubits holds up to three variables, in the correlators $\langle X_aX_b\rangle$, $\langle Y_aY_b\rangle$, and $\langle Z_aZ_b\rangle$ (30 variables); the OCSVM uses 16 qubits for 320 variables. (d) The PCE circuit: $R_Y$ and $R_Z$ rotations and $R_{ZZ}$ gates between neighboring qubits of the IBM chip (on the right, the couplings of the 16 qubits); the measured correlators feed training and decoding.}
\label{fig:overview}
\end{figure*}

This article lies at the intersection of predictive maintenance, machine learning, and quantum computing. We present a working pipeline for anomaly detection at the edge based on vibration analysis with the one-class support vector machine (one-class SVM, OCSVM) (Figure~\ref{fig:overview}a) and adopt quantum computing as the ``trainer'' of the pipeline. We evaluate the pipeline and the quantum algorithms on the public bearing dataset of Paderborn University.

The pipeline we developed takes as input a vibration signal with a high sampling rate and a duration of 1~s, and gives a binary classification as output, indicating whether the characteristics of the vibration are abnormal. The sampling rate of the dataset used in the tests is 64~kHz. The first step of the pipeline is feature extraction, which summarizes the signal of 64,000 samples in 5 numbers, and the second is classification with the OCSVM. As adaptations for the edge, we use a feature-selection process, in which a quantum optimizer chooses the 5 most suitable features out of a total of 16 known ones (which is necessary because computing each feature takes time), and we configure the OCSVM so that classification needs no floating-point operations. The quantum methods used are the variational quantum eigensolver for feature selection and variational quantum optimization with Pauli-correlation encoding for training the OCSVM.

The application of quantum computing to machine-learning and optimization problems is a relevant and challenging research topic. In recent years, major advances have started the noisy intermediate-scale quantum (NISQ) era \cite{preskill2018}, marked by working commercial quantum computers that are limited in the number and quality of their qubits. These limitations rule out the practical application of quantum algorithms with theoretically proven speedups, such as the algorithms of Shor \cite{shor1997} and Grover \cite{grover1996}. In machine learning and optimization, moreover, obtaining and demonstrating quantum advantage is hard even from a theoretical point of view, and many researchers regard the question as still open \cite{schuld2022}.

In our project, feature selection and the training of the OCSVM are modeled as quadratic unconstrained binary optimization (QUBO) problems, which are the specialty of quantum optimizers. The use of quantum optimization is justified by the adaptations for the edge: with binary weights, training the OCSVM, which is convex and solved in polynomial time in the usual case, becomes combinatorial, as does feature selection under a budget. The tests were run with few qubits on quantum simulators and with small instances of the problems (easy for classical computers), for demonstration purposes only, and random solutions are used as the baseline. The goal is not to demonstrate quantum advantage.

%==============================================================================
\section{THE PARTS OF THE PIPELINE}
%==============================================================================

\subsection{Feature Extraction}

A feature is a number computed from a 1-s window of the signal. We consider 16: ten time-domain statistics, computed from the raw signal, and six spectral descriptors, which need a fast Fourier transform (FFT), computed only once for all of them (Table~\ref{tab:features}).

\begin{table*}
\caption{The 16 features and their costs in thousands of cycles per window on the ESP32-C6, in fixed point. Notation: $x[n]$, $n = 1, \dots, N_s$, is the window ($N_s = 64{,}000$ samples), with mean $\bar{x}$, standard deviation $\sigma$, and peak $x_p = \max_n \lvert x[n] \rvert$; $\overline{\lvert x \rvert}$ and $\overline{\sqrt{\lvert x \rvert}}$ are the means of $\lvert x[n] \rvert$ and $\sqrt{\lvert x[n] \rvert}$; $X[q]$ is the magnitude spectrum at the frequencies $f_q$, and $p_q = X[q]^2 / \sum_{q'} X[q']^2$ the normalized power spectrum; $\mathcal{B} = [2, 20]$~kHz. The cost of a spectral feature excludes the FFT, paid once per window if any of them is used (last row).}
\label{tab:features}
\tablefont
\begin{tabular*}{\textwidth}{@{\extracolsep{\fill}}lllr@{}}
\toprule
Feature & Category & Formula & k cycles \\
\colrule
RMS & time & $\sqrt{\frac{1}{N_s} \sum_{n} x[n]^2}$ & 576 \\
Peak & time & $x_p = \max_n \lvert x[n] \rvert$ & 576 \\
Peak to peak & time & $\max_n x[n] - \min_n x[n]$ & 512 \\
Crest factor & time & $x_p / \mathrm{RMS}$ & 896 \\
Shape factor & time & $\mathrm{RMS} / \overline{\lvert x \rvert}$ & 960 \\
Impulse factor & time & $x_p / \overline{\lvert x \rvert}$ & 768 \\
Clearance factor & time & $x_p / \bigl(\overline{\sqrt{\lvert x \rvert}}\bigr)^2$ & 2,816 \\
Margin factor & time & $x_p / \mathrm{RMS}^2$ & 896 \\
Kurtosis & time & $\frac{1}{N_s} \sum_{n} \bigl(\frac{x[n]-\bar{x}}{\sigma}\bigr)^4$ & 1,601 \\
Skewness & time & $\frac{1}{N_s} \sum_{n} \bigl(\frac{x[n]-\bar{x}}{\sigma}\bigr)^3$ & 1,601 \\
Band energy & spectrum & $\sum_{f_q \in \mathcal{B}} X[q]^2$ & 129 \\
Spectral centroid & spectrum & $\mu_f = \sum_q f_q\, p_q$ & 393 \\
Spectral spread & spectrum & $\sigma_f = \sqrt{\sum_q (f_q - \mu_f)^2\, p_q}$ & 885 \\
Spectral skewness & spectrum & $\sum_q (f_q - \mu_f)^3\, p_q \,/\, \sigma_f^3$ & 1,279 \\
Spectral kurtosis & spectrum & $\sum_q (f_q - \mu_f)^4\, p_q \,/\, \sigma_f^4$ & 1,148 \\
Spectral entropy & spectrum & $-\sum_q p_q \ln p_q$ & 1,376 \\
\emph{FFT (shared stage)} & spectrum & mean removal, Hann window, 65,536-point real FFT, power spectrum & 13,576 \\
\botrule
\end{tabular*}
\end{table*}

Computing each feature has a cost, which we measure as the estimated number of machine cycles needed on the ESP32-C6. Computing all 16 features costs about \val{totalCycles} million cycles per window, of which the FFT accounts for \val{fftShare}. In this project, we assume a latency budget of at most \val{budget} million cycles per window, which makes it necessary to choose a subset of these features for the pipeline; the choice follows the classic mRMR criterion (maximum relevance, minimum redundancy) \cite{peng2005}. We choose 5 of the 16 features, which gives 4368 possible combinations.

\subsection{The One-Class Support Vector Machine}

The one-class support vector machine (OCSVM) \cite{scholkopf2001} is an anomaly detector that learns from examples of a single class only and classifies new examples as belonging to that class or not. This is useful in predictive maintenance, where there are many more real data from the healthy machine than labeled data from the machine with an anomaly. The OCSVM is trained only with windows of the signal of the healthy machine and learns the regions of normal behavior; a window that falls outside these regions is flagged as an anomaly.

To classify a vector $u$ holding the 5 features, the OCSVM computes the decision function $p(u)=\operatorname{sign}(f(u))$ from the score
\begin{equation}
f(u)= \sum_{i} \lambda_i K(v_i, u) - \rho,
\label{eq:score}
\end{equation}
where $u$ is an input vector, $v_i$ are the feature vectors of the training windows, $\lambda_i$ is a Lagrange multiplier (which we simply call a weight), $\rho$ is the classification threshold, and $K(\cdot,\cdot)$ is a kernel function. The weights are traditionally nonnegative real numbers, and the vectors $v_i$ for which $\lambda_i > 0$ are called support vectors. Only these vectors need to be stored by the program that performs the classification. The kernel works as a similarity measure and must satisfy $K(u,v)=\langle \phi(u), \phi(v) \rangle$, where $\phi(\cdot)$ is a linear or nonlinear feature map.

As adaptations for the edge, we adopt the base-2 Laplacian kernel of Anguita et al.\ \cite{anguita2006}, $K(\vec{u}, \vec{v}) = 2^{-\gamma \lVert \vec{u}-\vec{v} \rVert_1}$, which requires only fixed-point subtractions, additions, shifts, and table lookups, and we restrict the weights to $\lambda_i \in \{0, 1\}$, so that the weighted sum of Eq.~(\ref{eq:score}) becomes a plain sum. We set the number of support vectors to $w = 32$, chosen among the 320 healthy training windows.

Binarizing the weights creates a difficulty: the SMO algorithm \cite{platt1998}, traditionally used to train the OCSVM optimally in polynomial time, is incompatible with this adaptation, which makes training the OCSVM a combinatorial problem, NP-hard in general. This is overcome with classical or quantum combinatorial-optimization metaheuristics. Moreover, the usual rule $\rho=\sum_{i} \lambda_i K(v_i, v_j)$, computed with a support vector $v_j$ on the margin, can no longer be used to compute $\rho$, because with binary weights there are no support vectors on the margin; $\rho$ then requires a final supervised tuning step with data from artificial faults.

\subsection{Quantum Optimization and the QUBO}

Feature selection and the training of the adapted OCSVM are both written as QUBO problems \cite{lucas2014}, which consist of finding the list of 0/1 variables $\vec{x}$ that minimizes a cost
\begin{equation}
\vec{x}^* = \arg \min_{\vec{x}\in\{0,1\}^N} x^T Q x = \arg \min_{\vec{x}\in\{0,1\}^N} \sum_{i \leq j} Q_{i,j}\, x_i x_j.
\label{eq:qubo}
\end{equation}
The upper-triangular matrix $Q$ says how much each variable and each pair cost. Quantum optimizers turn this cost into the energy $E(x)$ of a system of qubits and search for the lowest-energy state, which encodes the solution \cite{kadowaki1998,farhi2014}. Since the QUBO has no constraints, the rules of the problem must enter the quantum circuit or the cost itself, as penalties.

\subsection{VQE With Hamming-Weight Preservation}

The variational quantum eigensolver (VQE) \cite{peruzzo2014} is a quantum-optimization metaheuristic of the family of variational quantum algorithms (VQAs). In VQAs, a parameterized quantum circuit, called the \emph{ansatz}, works as a sampler of solutions: each measurement produces a list of bits, and a classical optimizer of continuous variables, such as Adam or SPSA, tunes the parameters of the circuit so that good solutions are sampled more often. In the VQE, each variable of the QUBO is a qubit, and each computational-basis state $\ket{x_N\dots x_1}$ assigns a sequence of values to the variables $x_i$ of the QUBO. The quantum circuit is $U_\mathrm{ansatz}(\theta)U_\mathrm{prep}\ket{0 \cdots 0}$, where $U_\mathrm{prep}$ is a state-preparation operator and $\theta$ is the vector of parameters $\theta_i$ of the ansatz. The final state of the circuit is a superposition of candidate solutions, $\ket{\psi(\theta)} = \sum_i a_i(\theta) \ket{x_N\dots x_1}^{(i)}$, which, when measured in a shot, collapses into a solution state $\ket{x_N\dots x_1}^{(i)}$ with probability $|a_i(\theta)|^2$.

To tune $\theta$, the cost function of Eq.~(\ref{eq:qubo}) is rewritten as the eigenvalues of the Ising Hamiltonian $H$, that is, $H\ket{x} = \bigl(E(x) - C\bigr)\ket{x}$, where $E(x)=x^T Q x$ and $C$ is a constant, which does not change the lowest-energy state and is therefore omitted. The quantity that the optimizer minimizes is the mean energy $\bra{\psi(\theta)} H \ket{\psi(\theta)}$. The Hamiltonian $H$ is the linear operator
\begin{equation}
H = \sum_{i>j} J_{i,j} Z_i Z_j + \sum_i h_i Z_i,
\label{eq:ising}
\end{equation}
where $Z_i$ is the Pauli-$Z$ operator acting on the $i$th qubit, $J$ is the matrix of coupling coefficients, and $h$ is the vector of local coefficients. With the change of variables $x_i = (1 - z_i)/2$, where $z_i = \pm 1$ is the eigenvalue of $Z_i$, these coefficients are computed from $Q$ as
\begin{equation}
J_{i,j} = \frac{Q_{i,j} + Q_{j,i}}{4}, \qquad h_i = -\frac{Q_{i,i}}{2} - \frac{1}{4}\sum_{j \neq i} \left(Q_{i,j} + Q_{j,i}\right).
\label{eq:jh}
\end{equation}

In this work, we adopt a Hamming-weight-preserving circuit for the VQE (Figure~\ref{fig:overview}b), based on the pyramid of RBS gates of Landman et al.\ \cite{landman2022}. With this circuit, every solution state $\ket{x}^{(i)}$ has the same fixed number $k$ of ones and $n-k$ of zeros on the $n$ feature qubits, which is what fixes the number of selected features. Thus, the QUBO needs no penalty for this constraint, and the search space shrinks from $2^{16}$ to 4368 solutions. The circuit is built in three parts. First, $U_\mathrm{prep}$ applies $X$ gates to $k$ of the $n = 16$ feature qubits, preparing a state with exactly $k$ ones, such as $\ket{0\cdots0\,1\cdots1}$. Then, the ansatz applies only RBS (reconfigurable beam splitter) gates, each with its own angle $\theta$, to pairs of neighboring qubits. An RBS gate leaves $\ket{00}$ and $\ket{11}$ unchanged and only rotates the amplitude between $\ket{01}$ and $\ket{10}$:
\begin{equation}
\begin{aligned}
\mathrm{RBS}(\theta)\ket{01} &= \cos\theta\,\ket{01} - \sin\theta\,\ket{10},\\
\mathrm{RBS}(\theta)\ket{10} &= \sin\theta\,\ket{01} + \cos\theta\,\ket{10}.
\end{aligned}
\label{eq:rbs}
\end{equation}
It moves a one from a qubit to its neighbor but never creates or destroys a one; hence, any sequence of RBS gates preserves the Hamming weight. The gates are arranged in a pyramid: the pair of qubits $(i, i+1)$, with $i = 1, \dots, n-1$, gets $n - i$ gates, for a total of $n(n-1)/2$ gates (120 for 16 qubits) in $2n - 3$ layers. Finally, the last qubit, which represents the auxiliary variable $y_\mathrm{fft}$, explained below, and is not part of the Hamming-weight constraint, stays outside the pyramid and gets a single $R_Y(\phi)$ rotation, which leaves it in the state $\cos(\phi/2)\ket{0} + \sin(\phi/2)\ket{1}$.

We train the circuit in three ways: with Adam \cite{kingma2015}, using the exact energy and gradient (simulation); with SPSA \cite{spall1998}, which estimates a downhill direction from two energy estimates per iteration, made with 1024 shots, as on a real processor; and with COBYLA, a classical optimizer that uses no gradients. At the end, the circuit is measured 1024 times, and the answer is the lowest-cost measurement that meets the cycle budget, defined in the modeling section (the \emph{readout check}).

\subsection{PCE: Many Variables in Few Qubits}

Training SVMs with quantum optimization is not new: Date et al.\ \cite{date2025} demonstrate the use of quantum annealing (QA) to train the SVM, and the same formulation can easily be transferred to the VQE or to other quantum optimizers, such as QAOA \cite{farhi2014} and FALQON. However, with the VQE, QA, QAOA, or FALQON, at least one qubit would be needed for each training window, which is unfeasible in practical cases with NISQ hardware. As a solution, we propose a VQA with Pauli-correlation encoding (PCE) \cite{sciorilli2025}, which drastically reduces the number of qubits used in the optimization. In PCE, each variable $x_i$ is represented by a \emph{correlation} between several qubits, two in our case, that is, by the expectation value $\langle Z_aZ_b\rangle$, $\langle X_aX_b\rangle$, or $\langle Y_aY_b\rangle$ of the pair of qubits $a$ and $b$, a real value between $-1$ and 1. The variable is computed as
\begin{equation}
x_i = \frac{1 + \operatorname{sgn}\langle \Pi_i \rangle}{2},
\label{eq:decode}
\end{equation}
where $\Pi_i \in \{X_aX_b, Y_aY_b, Z_aZ_b\}$.

With $n$ qubits, there are $n(n-1)/2$ pairs, and each pair offers three correlators ($X$, $Y$, and $Z$). Thus, $n$ qubits hold up to $3n(n-1)/2$ variables, and 320 variables fit in 16 qubits. The circuit (Figure~\ref{fig:overview}d) uses the native gates of IBM processors, between neighboring qubits of the chip; it was chosen for its cost of execution on this hardware, not for the structure of the problem. Training replaces the sign with a smooth version, $\tilde{x}_i = [1 + \tanh(\alpha\langle\Pi_i\rangle)]/2$, where $\alpha$ sets the steepness, and minimizes the cost computed with these variables,
\begin{multline}
\mathcal{L}(\vec{\theta}) = \sum_{i} Q_{i,i}\, \tilde{x}_i + \sum_{i<j} Q_{i,j}\, \tilde{x}_i \tilde{x}_j \\
+ \beta \nu \left[ \frac{1}{N} \sum_{i} \tanh^2(\alpha\langle\Pi_i\rangle) \right]^2,
\label{eq:loss}
\end{multline}
where $N$ is the number of variables, $\nu$ is a scale of the cost, and the last term, with weight $\beta$, is a regularization that keeps the correlators small. At the end, the correlators are measured and the $w$ largest become ones, $w$ being the preset number of support vectors. Then, we apply one round of classical local search, as in the work of the PCE authors, which swaps a one and a zero whenever this lowers the cost; results are evaluated before and after it.

Training the OCSVM with binary weights also requires fixing the Hamming weight, but doing so at the level of the quantum circuit is an open problem in PCE. The usual way, a penalty that grows when the sum of the smooth variables moves away from $w$, fails when the choice is sparse: it asks for almost all correlators to be strongly negative, which is impossible, because, for each Pauli axis $P$, $\sum_{a<b}\langle P_aP_b\rangle \geq -n/2$. As a palliative solution, we use an \emph{offset} term $b$, adjusted at every step so that the smooth variables add up to exactly $w$,
\begin{equation}
\tilde{x}_i = \frac{1 + \tanh\bigl(\alpha(\langle\Pi_i\rangle - b)\bigr)}{2}, \quad b \text{ such that } \sum_{i} \tilde{x}_i = w.
\label{eq:offset}
\end{equation}
We train PCE with Adam (exact simulation) and with SPSA (1024 shots).

%==============================================================================
\section{MODELING THE TWO PROBLEMS}
%==============================================================================

\subsection{Feature Selection}

With $x_i = 1$ when feature $i$ is chosen, the cost is mRMR, the mean redundancy minus the mean relevance of the subset:
\begin{multline}
f(\vec{x}) = -\frac{1}{5}\sum_i a_i x_i + \binom{5}{2}^{-1} \sum_{i<j} B_{i,j}\, x_i x_j, \\
\sum_i x_i = 5, \qquad \sum_i c_i x_i + C_{\mathrm{fft}}\, y \leq C_{\max}.
\label{eq:fs}
\end{multline}
$a_i$ is the relevance of feature $i$, the mutual information \cite{kraskov2004} mapped to $[0, 1]$; $B_{i,j}$ is the absolute correlation between features $i$ and $j$; $c_i$ is the number of cycles of feature $i$, $C_{\mathrm{fft}}$ that of the FFT, and $C_{\max}$ the budget of \val{budget} million, which about \val{feasibleShare} of the subsets meet. The FFT is paid once if \emph{any} spectral feature is used, which is not quadratic; so we add an auxiliary variable $y_{\mathrm{fft}} = y$, meaning ``the FFT is computed,'' and a small penalty that makes it always worse to use a spectral feature with $y = 0$. The first constraint lives in the circuit. The second, the budget, is an inequality and does not fit in the circuit; it enters through the unbalanced penalization \cite{montanez2024}, which avoids the slack variables (and the extra qubits) of the usual approach: the cost gets the term $-\eta_1 s + \eta_2 s^2$, with $s = (C_{\max} - \sum_i c_i x_i - C_{\mathrm{fft}}\, y)/C_{\mathrm{total}}$ the normalized leftover budget and $C_{\mathrm{total}}$ the cost of the 16 features. The parabola grows quickly when the budget is exceeded ($s < 0$) and weighs little when it is met with a small leftover; it is an approximation, because it also grows for large leftovers, and every answer is therefore checked at readout. The weights $\eta_1$ and $\eta_2$ were chosen on a grid, checking by enumeration that the constrained optimum is the feasible subset of lowest energy and that the fewest possible subsets over the budget (\val{infeasibleBelow}, in all validation rounds) lie below it. There are 17 variables: the VQE uses 16 qubits with exactly 5 ones for the features and a free qubit for $y$ (Figure~\ref{fig:overview}b).

\subsection{OCSVM}

With binary multipliers, training the OCSVM reduces to
\begin{equation}
\min_{\vec{\lambda}} \sum_{i<j} K(v_i, v_j)\, \lambda_i \lambda_j \quad \text{with} \quad \sum_i \lambda_i = w,
\label{eq:ocsvm}
\end{equation}
that is, choosing the $w$ healthy windows least similar to each other, a set of diverse prototypes of normal operation. Relaxing the weights to $0 \leq \lambda_i \leq 1$ gives back the convex problem of the usual OCSVM (with $\nu = w/320$), which SMO solves exactly and whose optimum bounds the binary cost from below. There are 320 variables with 32 ones, which PCE places in 16 qubits. A new window gets the score $f(u)$ of Eq.~(\ref{eq:score}) and is deemed faulty if $f(u) < 0$, that is, if the sum of its similarities to the prototypes falls below the threshold $\rho$.

%==============================================================================
\section{HOW WE VALIDATED THE PIPELINE}
%==============================================================================

\textbf{Data and experiments.} We use the Paderborn University bearing dataset \cite{lessmeier2016}, with 6 healthy bearings, 12 with artificial damage, and 14 with real damage, cutting the vibration into 1-s windows. The experiment is repeated in three \emph{rounds}: in each one, two healthy bearings are kept for testing, together with the 14 with real damage, and the other four (320 windows) are used for training. Each round thus has its own feature selection and its own training of the OCSVM.

In each round, each optimizer is run several times (\val{vqeSeeds} in the VQE and 5 in PCE), each time starting from randomly drawn initial parameters, and, for comparison, we draw random solutions (\val{randomDraws} feature subsets and \val{ocRandom} choices of support vectors). The figures pool the runs and draws of the three rounds.

\textbf{Supervised and unsupervised parts.} The OCSVM learns in an unsupervised way from healthy windows only, but two steps of the training of the pipeline use data from artificial-damage bearings: the relevance of the features and the decision threshold $\rho$. The threshold is chosen in a supervised way to maximize the balanced accuracy. Selecting features and computing the threshold in an unsupervised way was left as an open problem in this study.

\textbf{Comparisons and metrics.} In feature selection, each answer is compared with the exact optimum, obtained by enumeration, through the normalized gap (0 is the optimum and 1 the mean of a random choice) and through the F1 score of a conventional OCSVM (scikit-learn \cite{pedregosa2011}) trained with the chosen features. For the binary OCSVM, the reference is the best of 10 classical local searches; besides random choices, a classical baseline rounds to 1 the 32 largest weights of the OCSVM trained by SMO (SMO + rounding). The detectors are evaluated on the real damage by accuracy, F1, and AUC (the probability that a faulty window gets a higher anomaly score than a healthy one).

\textbf{Cost model.} We count the operations of every loop of every feature, with typical costs of a RISC-V core without floating point, and calibrate the FFT on the published timings of the ESP-DSP library \cite{espdsp}; it is an analytical estimate, not a measurement on the device.

%==============================================================================
\section{RESULTS}
%==============================================================================

\subsection{Feature Selection With the VQE}

\begin{figure*}
\centerline{\resultfigure{\textwidth}{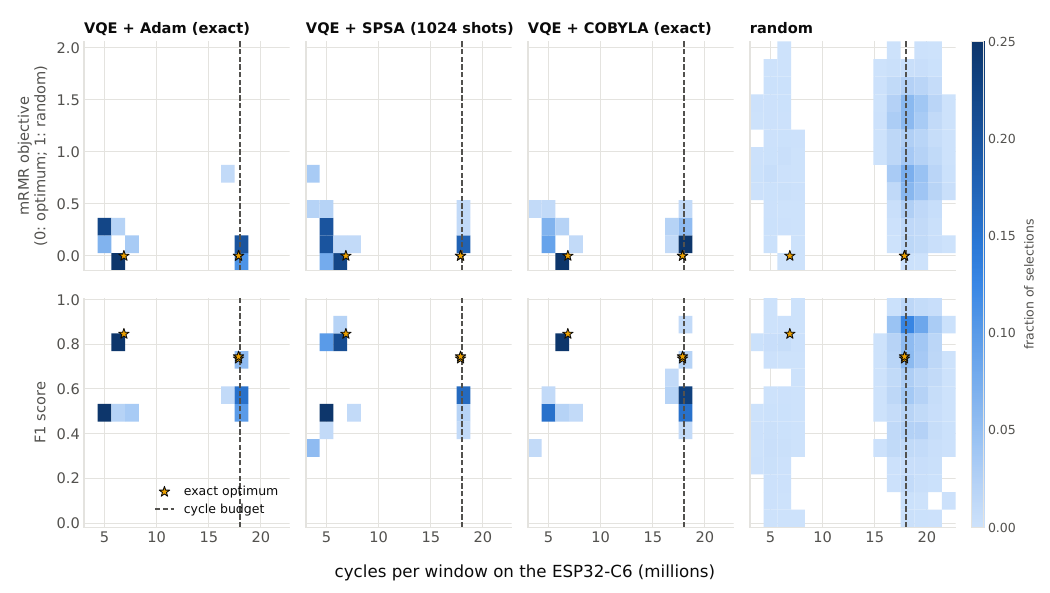}}
\caption{2D histograms of the VQE solutions, with Adam, SPSA, and COBYLA, and of random solutions in the selection of 5 of 16 features. The vertical axes show, in the top row, the value of the normalized mRMR cost function (0: exact optimum within the budget; 1: mean of a random choice) and, in the bottom row, the F1 of the conventional OCSVM (scikit-learn) with the chosen features. The horizontal axis shows the number of cycles needed to compute the features on the ESP32-C6, with the budget (upper limit) dashed. The stars are the exact optima of each round of the experiment within the space of valid solutions.}
\label{fig:selection}
\end{figure*}

In the \val{vqeRuns} runs of each optimizer, the VQE found the optimal subset in \val{optAdam} of the cases with Adam, \val{optCobyla} with COBYLA, and \val{optSpsa} with SPSA, with mean gaps to the optimum of \val{gapAdam}, \val{gapCobyla}, and \val{gapSpsa} (Figure~\ref{fig:selection}, top). The cycle budget was met by all answers of COBYLA and SPSA and by \val{feasAdam} of those of Adam; random choices met it in \val{feasRandom} of the cases.

For detection (Figure~\ref{fig:selection}, bottom), the constrained optimum had the highest mean F1, \val{f1Exact}, against \val{f1Random} for the random choices. The answers of the VQE that missed the optimum, however, often traded the FFT for cheap time-domain features and had a lower F1: closeness in the mRMR objective did not always translate into good detection.

\subsection{Training the OCSVM With PCE}

\begin{figure}
\centerline{\resultfigure{\columnwidth}{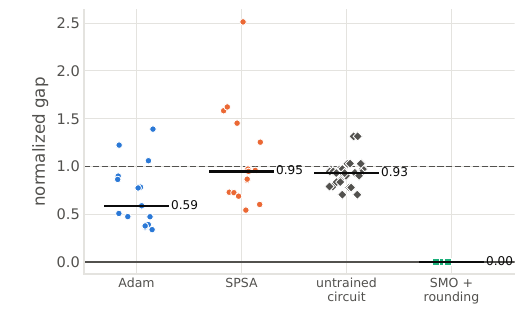}}
\caption{Spread of the gaps of the decoded OCSVM solution, before local search, to the best known solution (0), for PCE with the offset, trained by Adam and by SPSA, and for the classical SMO + rounding baseline (one point per round). The dashed line is the level of a random choice, and ``untrained'' is the circuit with its initial parameters. Black bars: medians.}
\label{fig:constraint}
\end{figure}

Figure~\ref{fig:constraint} shows the quality of the decoded PCE solutions, before local search. With the offset, PCE + Adam reached a median gap of \val{ocRawGapOffset} to the best known solution, below the untrained circuit (\val{ocRawGapUntrained}); with SPSA, it stayed at \val{ocRawGapSpsa}. In a first version of this study, with the usual penalty, this gap was \val{ocRawGapPenalty}: the circuit learned almost nothing.

\begin{figure*}
\centerline{\resultfigure{\textwidth}{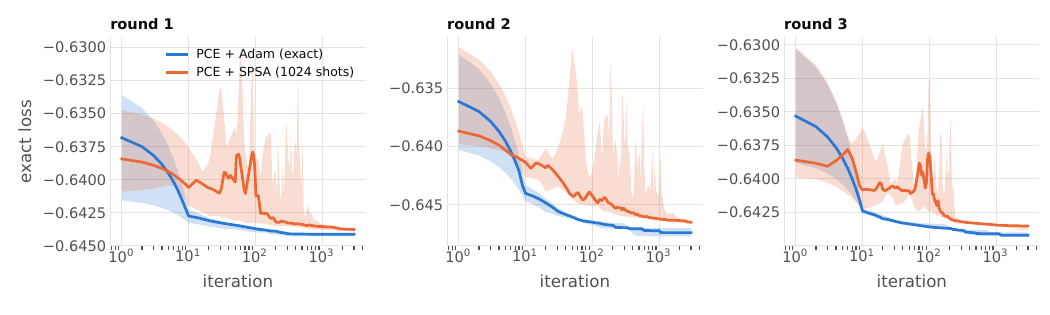}}
\caption{Exact loss~(\ref{eq:loss}) during the training of PCE on the OCSVM (16 qubits, 320 variables), in each round, for Adam with exact simulation and for SPSA with the loss estimated from 1024 shots (the exact loss of the current parameters was computed only for the plot). Line: median of the 5 runs; band: minimum and maximum. An Adam run that stopped by the convergence criterion keeps its last value. Logarithmic horizontal axis.}
\label{fig:loss}
\end{figure*}

Adam converged smoothly, in a median of \val{ocAdamSteps} steps; SPSA oscillated, used all \val{ocSpsaIters} iterations (\val{ocSpsaShots} shots per run), and ended slightly above Adam (Figure~\ref{fig:loss}).

\begin{figure*}
\centerline{\resultfigure{\textwidth}{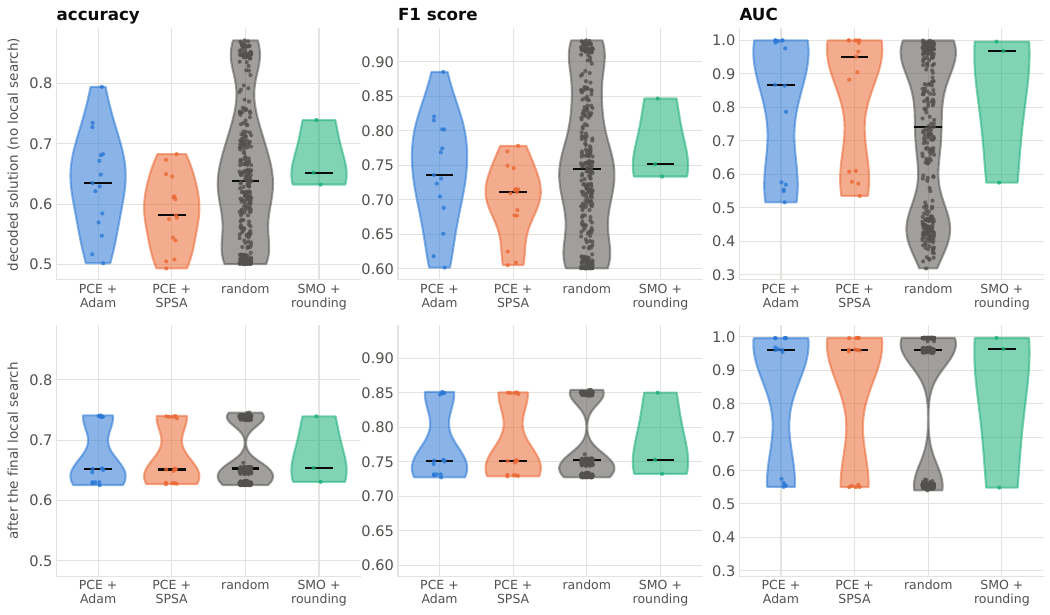}}
\caption{Accuracy, F1, and AUC on the real damage of the detectors with binary multipliers trained by PCE (Adam with exact simulation; SPSA with 1024 shots), by random choices of 32 windows, and by SMO + rounding (one point per round), before (top) and after (bottom) the same round of local search; one point per run or draw, bar at the median.}
\label{fig:violins}
\end{figure*}

Before local search, both PCE arms gave detectors with a higher mean AUC than those of random choices of 32 windows (\val{ocAucAdamDec} and \val{ocAucSpsaDec}, against \val{ocAucRandDec}), but not a higher F1 (Figure~\ref{fig:violins}, top). After one round of local search, the three methods reached the same detectors, with an accuracy of \val{ocAccAdamSrch}, an F1 of \val{ocF1AdamSrch}, and an AUC of about \val{ocAucAdamSrch} (Figure~\ref{fig:violins}, bottom): at this size, the classical search alone brings any starting point close to the best known solution.

The SMO baseline shows how easy this instance is. Only \val{smoSupport} of the 320 weights of the relaxation were nonzero, \val{smoBounded} of them already at the upper bound, and its lower bound was within \val{smoLbGap} of the best known solution. Rounding gave gaps of at most \val{smoRoundGapMax} (Figure~\ref{fig:constraint}) and the same detectors as the other methods (Figure~\ref{fig:violins}).

%==============================================================================
\section{CONCLUSION}
%==============================================================================

We showed how to take constrained ``choose a few of many'' problems to quantum optimizers, on a real case of predictive maintenance at the edge. When the problem is small, the VQE with gates that preserve the number of ones imposes the cardinality through the circuit itself. When the problem is large, PCE stores hundreds of variables in few qubits, but the constraints must be imposed classically during training. In both cases, the optimizer must be compared with random choices and with classical baselines, and the answer assessed with the metric of the application: here, the convex relaxation of the OCSVM left only \val{smoFree} fractional weights, a residual problem that a few qubits, or enumeration, can solve, which suggests relaxing classically first and optimizing only that residue.

The study uses a single operating condition, six healthy bearings from one manufacturer, simulated circuits, and an analytical cost model. The relevance of the features and the decision threshold still depend on artificial-damage bearings; computing the threshold in an unsupervised way remains an open problem. The PCE circuit followed the native gates of the hardware; we suggest, as a topic for future research, investigating PCE circuits designed for SVM training, which exploit the theoretical structure of the problem, such as its symmetries and the sparsity of the solution, instead of merely following a pattern of native gates and maximal expressivity. Next steps include running the circuits on hardware, measuring the cycles on a real ESP32-C6, applying the techniques to larger problems, and comparing them with stronger baselines.

%==============================================================================
\section{CODE AND DATA AVAILABILITY}
%==============================================================================

The code, executable with a single command, and the instructions for reproducing the experiments are available at \url{https://github.com/SIMPred-Petrobras/pce-vqe-pdm-paper-code}. The Paderborn University bearing dataset is publicly available (license CC BY-NC 4.0).

% TODO: confirm the address and that the repository will be public at
% submission time.

%==============================================================================
\section{ACKNOWLEDGMENTS}
%==============================================================================

This research was carried out within the SimPred project, conducted by the Laboratory of Applied Computational Intelligence (LICA) of the Federal University of Alagoas (UFAL) in partnership with Petrobras. The authors, who hold research fellowships from the project, thank Petrobras for its financial support and the LICA/UFAL team.

\def\refname{REFERENCES}

\bibliographystyle{IEEEtran}
{\small\bibliography{references}}

\begin{authorbio}{Filipe Chagas Ferraz}is a master's student in computer science at the Federal University of S\~ao Carlos, S\~ao Carlos, SP, 13565-905, Brazil, an external researcher in the SimPred project, and a collaborator at Dobslit, a Brazilian quantum technology startup. Ferraz received a B.S. degree in computer engineering from the Federal University of Mato Grosso. Contact Ferraz at \href{mailto:filipe.ferraz@estudante.ufscar.br}{filipe.ferraz@estudante.ufscar.br}.
\end{authorbio}

\end{document}